\documentclass[journal]{IEEEtran}
\usepackage{graphicx, caption}
\usepackage{authblk}
\usepackage[utf8]{inputenc}
\usepackage{gensymb,textcomp}
\usepackage{booktabs,siunitx}
\usepackage{indentfirst}
\usepackage{footnote}
\usepackage{threeparttable}

\usepackage{amsmath,amsfonts,amssymb,amsthm}
\usepackage{mathtools}
\usepackage{commath}
\usepackage[sc,osf]{mathpazo}
\usepackage{booktabs}
\usepackage{multirow}
\usepackage{dirtytalk}
\usepackage{makecell}
\usepackage{graphicx}
\usepackage[sorting=none]{biblatex}
\usepackage{threeparttable}
\usepackage{float}
\usepackage{subcaption}
\newcommand{\etal}{\textit{et al}.}

\begin{document}
\title{Late fusion of machine learning models using passively captured interpersonal social interactions and motion from smartphones predicts decompensation in heart failure}

\author{Ayse~S.~Cakmak,
        Samuel~Densen,
        Gabriel~Najarro,
        Pratik~Rout,
        Christopher~J.~Rozell,
        Omer~T.~Inan, 
        Amit~J.~Shah,
        and~Gari~D.~Clifford~\IEEEmembership{Senior Member,~IEEE }
\thanks{Ayse S. Cakmak, Christopher J. Rozell and Omer T. Inan are with the Department of Electrical and Computer Engineering, Georgia Institute of Technology, Atlanta, GA, 30332 USA. Gari D. Clifford is with The Wallace H. Coulter Department of Biomedical Engineering, Georgia Institute of Technology, Atlanta, GA, 30332 USA and the Department of Biomedical Informatics, School of Medicine, Emory University, Atlanta, GA, 30322 USA. Amit J. Shah and Pratik Rout are with the Department of Epidemiology, Rollins School of Public Health, Emory University, Atlanta, GA, 30322 USA. Samuel Densen is with School of Medicine, Emory University, Atlanta, GA, 30322 USA. Gabriel Najarro is with Emory Healthcare, Emory University, Atlanta, GA, 30322 USA.}
}

\maketitle

\begin{abstract} 
\textit{Objective:} Worldwide, heart failure (HF) is a major cause of morbidity and mortality and one of the leading causes of hospitalization. Early detection of HF symptoms and pro-active management may reduce adverse events. \textit{Approach:} Twenty-eight participants were monitored using a smartphone app after discharge from hospitals, and each clinical event during the enrollment (N=110 clinical events) was recorded. Motion, social, location, and clinical survey data collected via the smartphone-based monitoring system were used to develop and validate an algorithm for predicting or classifying HF decompensation events (hospitalizations or clinic visit) versus clinic monitoring visits in which they were determined to be compensated or stable. Models based on single modality as well as early and late fusion approaches combining patient-reported outcomes and passive smartphone data were evaluated. \textit{Results:} The highest AUCPr for classifying decompensation with a late fusion approach was 0.80 using leave one subject out cross-validation. \textit{Significance:} Passively collected data from smartphones, especially when combined with weekly patient-reported outcomes, may reflect behavioral and physiological changes due to HF and thus could enable prediction of HF decompensation.
\end{abstract}


\IEEEpeerreviewmaketitle

\section{Introduction}
\IEEEPARstart{T}{he} American Heart Association estimates that between 2013 and 2016, approximately 6.2 million Americans had heart failure (HF), an increase of 20\% over the four years prior \cite{virani2020heart}. In 2012, the economic burden of HF was estimated at \$30.7 billion. Projections suggest a 127\% increase in cost by 2030. Overall, cardiovascular diseases account for the highest expenditures amongst all non-communicable diseases in the US \cite{chen2018macroeconomic}. 

HF decompensation, associated with hypervolemia (volume overload), is defined as a clinical syndrome in which a functional change in the heart leads to new or increasing symptoms, including fatigue, dyspnea, and edema, and requires hospitalization \cite{felker2003problem}. Treatment includes diuretics and vasodilators intended to improve volume status and cardiac function. Unfortunately, even following successful treatment and return to the euvolemic (normal volume status) state, decompensation episodes can continue to occur with increasing frequency \cite{felker2003problem, joseph2009acute}. Patil \etal reported that about 20\% of the patient cohort were readmitted within 30 days of initial hospitalization due to HF, with a median readmission time of 12 days \cite{patil2019readmissions}. Furthermore, patients with a lower income had a higher readmission rate, indicating that socio-economical factors could also contribute to the disease's progression. If inexpensive monitoring approaches are developed to identify decompensation episodes developing outside the clinic, medical interventions could then be administered proactively to prevent hospitalization or other adverse outcomes. 

Various studies have investigated techniques for monitoring HF patients non-intrusively. Packer \etal \cite{packer2006utility} showed that using a combination of clinical variables and impedance cardiography features could be a predictor of a decompensation event in the next 14 days. Previous studies have also investigated the use of wearable devices adhered to the chest. In the `Multisensor Monitoring in Congestive Heart Failure' study, the authors propose an algorithm that uses physiological signals, and they report a sensitivity of 63\%, and specificity of 92\% \cite{anand2012design}. However, the authors provide few details and claim it is `proprietary'. Inan \etal recorded seismocardiogram signal with a non-invasive wearable patch before and after a 6-minute walk test to analyze the cardiac response to exercise \cite{inan2018novel}. The authors used graph similarity scores between the rest and recovery phases and found a significant difference between compensated and decompensated groups. In another example, similarity-based modeling was used with physiological signals from a patch on the chest to detect changes from the baseline. This algorithm had a sensitivity of 88\% and specificity of 85\% \cite{stehlik2020continuous}. Using ballistocardiogram data recorded at home was also investigated \cite{aydemir2019classification}, and authors demonstrated that collecting high-quality ballistocardiogram data at home is feasible, and an AUC of 0.78 could be achieved for classifying clinical status. Other non-invasive approaches include patient-reported outcomes, which could be collected using clinically validated questionnaires such as Kansas City Cardiomyopathy Questionnaire (KCCQ). The KCCQ assesses the quality of life, predict readmissions and mortality in HF patients \cite{green2000development}. In a previous study, Flynn \etal reported that KCCQ had modest correlations with exercise capacity measured by the 6-minute walk test in a population with HF \cite{flynn2009relationships}.

With the advancement of technology, smartphones have become a ubiquitous part of our daily life. For long-term monitoring, using a smartphone could be advantageous to a solution requiring an additional device by reducing the disruption to patients' normal daily routine. Our research team and collaborators have previously developed the 'Automated Monitoring of Symptom Severity' (AMoSS) app, which is a custom and scalable smartphone-based framework for remote monitoring \cite{palmius2014multi}. Subsequently,  the current authors used the passive data from the first ten participants of this study to estimate the KCCQ surveys collected through the app \cite{cakmak2018personalized}. The model estimated the KCCQ score with a mean absolute error of $5.7\%$, providing an entirely passive method of monitoring HF related quality of life. (The method was passive in the sense that it does not require any active participation by either the patient or clinical staff beyond the everyday use of a mobile phone to monitor activity and behavioral patterns in the background using software.) Then, in subsequent work, motion data was used to classify decompensation or compensation events \cite{cakmak2019passive}. By using a hold-out test randomly sampled from $30\%$ of the events ($N_{test}=32$), the AUC of the classifier was found to be 0.76. 

In this work, heart failure decompensation events are predicted from features derived from passive and active data collected by the smartphone-based framework. Features were extracted from motion, social contact, location, and clinical survey data (KCCQ). Algorithms based on using a single modality and two different sensor fusion approaches were developed. An analysis of the feature importance in the model is also presented. Finally, a novel late-fusion model that combines the KCCQ, motion, and social contact data is proposed.

\section{Methods}
\subsection{Study overview and data collection}
Earlier research with the AMoSS app \cite{palmius2014multi} was augmented for use in this study. The app passively collected 3-D accelerometer data at 5Hz sampling frequency, location, clinical surveys, and digital social contact data. All data were de-identified at the source with hashed identifiers and random geographic offsets were added to the location data to protect the participant's privacy. The data was stored in Amazon Web Services data buckets and the app uploaded data every few hours.

\begin{figure}[!htbp]
    \centering
    \includegraphics[width=0.43\textwidth]{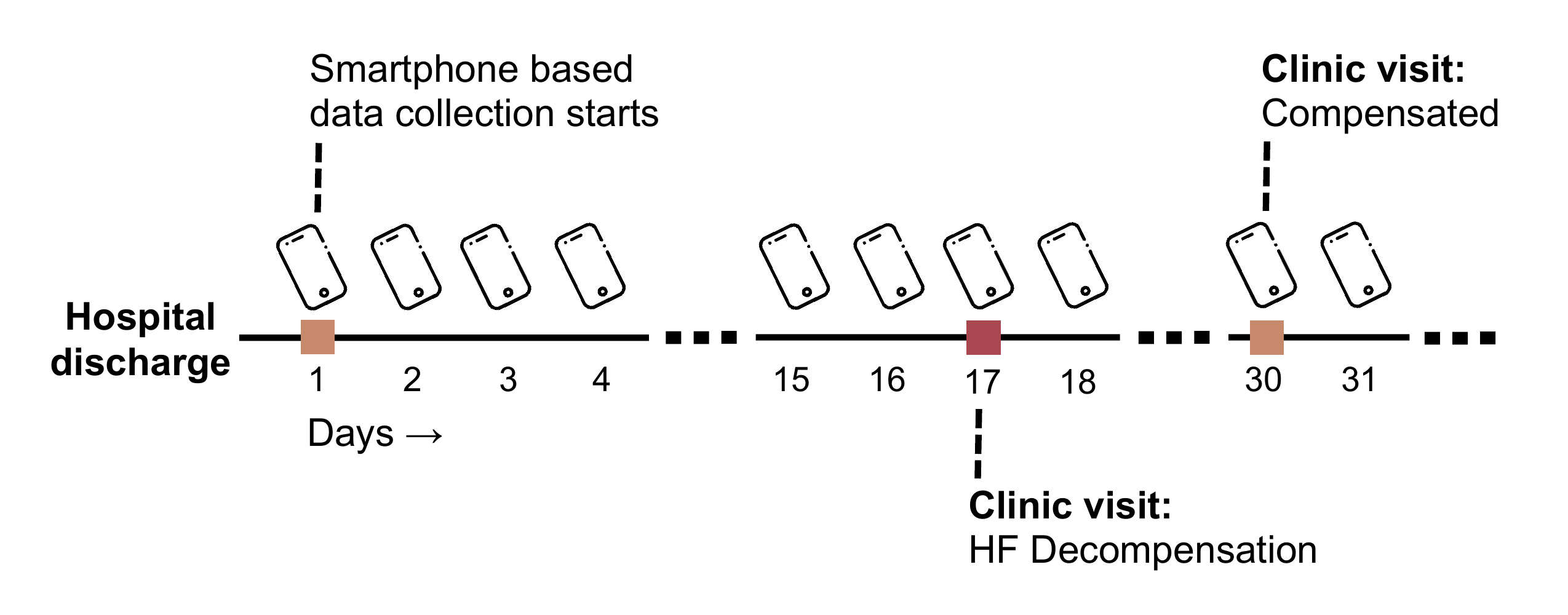}
    \caption{Illustration of the study timeline. Passive data collection started after the hospital discharge, and the clinical team recorded the clinical events after the enrollment.}
    \label{timeline}
\end{figure}

Participants with HF enrolled in the ongoing study at the Veterans Affairs Medical Center and Emory University Hospital in Atlanta, USA. The study protocol was approved by the IRB (\#00075867) at Emory University. The clinical team provided participants with an Android-based smartphone with the app installed during the enrollment. The participant could elect to stop sharing any data type during the study, using the switches provided in the app. Fig. \ref{timeline} illustrates the study timeline after the participant is enrolled. The app passively collected data while the clinical team recorded the clinical events, which consisted of hospital visits with compensated or decompensated status during the enrollment. 

The data from 28 participants (26 males and two females) who contributed at least one clinical event were used in this research. The inclusion criteria for the study were the following: participants needed to have a diagnosis consistent with congestive heart failure as noted in the electronic medical records within the Emory Health Network, be over the age of 18, able to consent to a clinical study and speak English as their primary language. Exclusion criteria were: diagnosis with a terminal illness with a life expectancy of fewer than six months, if they were enrolled in a hospice program, or enrollment in a clinical study that precluded them from participating in another clinical study. Finally, participants had to be willing and able to comply with the use of their smartphones, as indicated in the study.  Table \ref{dataset} shows more details about the participants in the dataset.

\begin{table}[htpb!]
\caption{Dataset description. If the metric is not available, the participant is excluded from that row.}
\centering
\label{dataset}
\begin{tabular}{@{}ll@{}}
\toprule[\heavyrulewidth]\toprule[\heavyrulewidth]
\vspace{0.1cm}
Num. comp. events & 62 \\
\vspace{0.1cm}
Num. decomp. events & 48 \\
\vspace{0.1cm}
Avg. comp. events per person & 2 \\
\vspace{0.1cm}
Avg. decomp. events per person & 2 \\
\vspace{0.1cm}
Avg. ejection fraction (EF) (\%) & 35 \\
\midrule
\vspace{0.1cm}
Gender & 93\% male \\
\vspace{0.1cm}
Age (mean $\pm$ std) & 67 $\pm$ 8 \\
\vspace{0.1cm}
BMI (mean $\pm$ std) & 31 $\pm$ 6 \\
Employment & \begin{tabular}[c]{@{}l@{}}Employed: 3\\ Unemployed: 5\\ Retired: 7\end{tabular} \\ \bottomrule
\end{tabular}
\end{table}

\subsection{Clinical events}
Clinical events consisted of decompensated and compensated events and were collected by the clinical team when the participants visited the hospitals. In the compensated events, the participants visited the hospital for any reason, and their fluid levels were determined to be normal based on the clinician assessment, which includes a history and physical examination. For the decompensated events, the clinical team determined the participant to have functional limitations related to HF. Decompensated and compensated events were assigned to positive and negative classes, respectively.

\subsection{Passive data sources}
\begin{figure}[!htbp]
    \centering
    \includegraphics[width=0.48\textwidth]{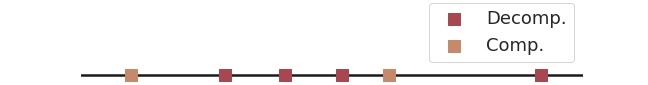}
    \includegraphics[width=0.48\textwidth]{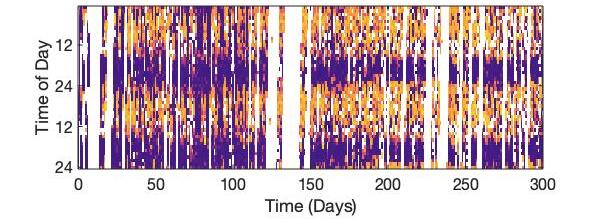}
    \caption{Double plot representation of actigraphy data, which illustrates daily motion intensity levels for one participant.  Darker colors indicate lower intensity movement, and the white color indicates missing data. On the top of the plot, decompensated and compensated clinical events are shown with red and orange squares respectively.}
    \label{doublePlot}
\end{figure}

The raw 3D accelerometer data was converted to activity counts using the Actigraphy Toolbox to reduce the required memory for storing \cite{actToolbox}. In the first step, the z-axis of the accelerometer data was filtered using a bandpass Butterworth filter with $0.25-11$ Hz passband to eliminate extremely slow or fast movements \cite{Ancoli-Israel2003}. Then, the maximum values inside 1-second windows were summed for each 30-second epoch to obtain the activity counts, following the approach described by Borazio \etal \cite{Borazio2014}. For this data type, if the participant shared data for less than 0.1\% of the analysis window, that window was considered missing. A common way for visualizing motion data in sleep studies to emphasize shifts in sleep rhythms is in the ``double plot" format, as shown in Fig. \ref{doublePlot}. This figure illustrates the motion data for one participant over a recording period of 300 days, and the darker colors indicate lower-intensity movement. Each column consists of two consecutive days of data stacked together. The first column shows motion intensity levels on days 1-2, and the second column shows days 2-3, and so on. White regions indicate missing data, which could be due to the participant turning off the data sharing or the smartphone running out of battery.

\begin{figure}[!htbp]
    \centering
    \includegraphics[width=0.48\textwidth]{figures/events_cut.jpg}
    \includegraphics[width=0.48\textwidth]{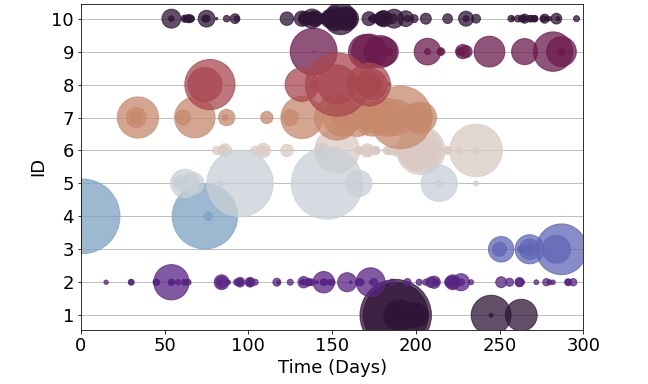}
    \caption{Participant's social contact intensity over 300 days. Each unique contact is assigned a number as shown in the y-axis, and the circle radius is proportional to call duration to each ID. On the top of the plot, decompensated and compensated clinical events are shown with red and orange squares respectively.}
    \label{socialContact}
\end{figure}

\begin{figure}[htbp!]
    \centering
    \includegraphics[width=0.35\textwidth]{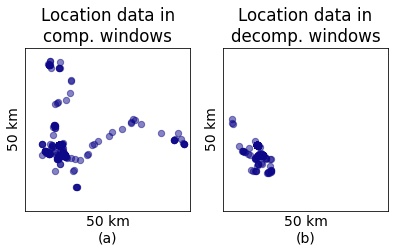}
    \caption{Location data collected in compensated and decompensated windows for a participant, shown on the same map with 50x50 km dimensions.}
    \label{locationPlot}
\end{figure}

\begin{figure}[htbp!]
    \centering
    \includegraphics[width=0.3\textwidth]{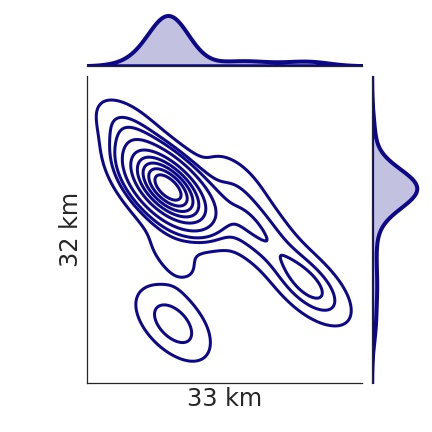}
    \caption{Kernel density estimate for the location data of one participant.}
    \label{locationKDE}
\end{figure}

Social contact data included the call data and the duration of each call. Each contact was anonymized and assigned a unique identifier at the source. Fig. \ref{socialContact} illustrates one participant's social contact over 300 days for the ten most frequently contacted IDs. Lastly, location data was collected using the Android location services application program interface, which generally used cellphone tower or WiFi and not GPS for geolocation. Fig. \ref{locationPlot} shows the location data of a participant, collected from compensated and decompensated windows. High spatial resolution was not required since the aim was to identify the general environment in which a user was located. (E.g., home, work, shops, etc.)  If the smartphone moved at least $100$ meters, and at least $5$ minutes had passed since the last location data update, a new relative location was recorded. These parameters were defined while designing the app to preserve battery life while still providing sufficient temporal and spatial resolution in comparison to the phone's ability to geo-locate without GPS. Fig. \ref{locationKDE} shows the kernel density estimate of one participant's all location data updates.

\subsection{Active data sources}
\begin{figure}[h!]
    \centering
    \includegraphics[width=0.48\textwidth]{figures/events_cut.jpg}
    \includegraphics[width=0.48\textwidth]{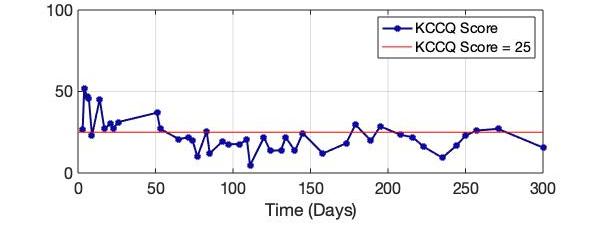}
    \caption{KCCQ summary score over days for the participant. KCCQ score $\leq25$ indicates a transition to severe HF. Decompensated and compensated clinical events are shown with red and orange squares respectively, above the plot.}
    \label{kccqPlot}
\end{figure}

The active data type, which required user input, was KCCQ administrated through the smartphone app. The scores are lower for severe HF symptoms, and KCCQ scores $\leq25$ correspond to New York Heart Association (NYHA) class IV.  In this study, we used the shorter version of the questionnaire, referred to as KCCQ-12 \cite{Jones2013}. The KCCQ-12 survey had physical limitation, symptom frequency, quality of life, and social limitation domains, and the summary score (ranging from 0-100) was the average of all domains. Fig. \ref{kccqPlot} shows the KCCQ-12 scores administrated through the app.

\subsection{Feature extraction from windows of data}
Several features were extracted from the data collected through the app to construct the motion feature set. A window of data was the N day period before a clinical event, and the feature extraction was performed for each window. The window size N was chosen to be 14 days initially since it was also selected by the developers of KCCQ to represent the participant's recent functioning \cite{green2000development}. Firstly, from preprocessed smartphone activity counts, descriptive statistics were extracted. These included mean ($act_{mean}$), standard deviation ($act_{std}$), mode ($act_{mode}$), skewness ($act_{skew}$), and kurtosis ($act_{kurt}$). The completeness percentage ($act_{comp}$) was calculated by dividing the epochs with data by the total number of epochs in the window. 

For each window, the total number of calls ($numCalls$), the sum of the duration of calls ($durCalls$), the standard deviation of the duration of calls ($durCalls_{std}$), the sum of durations without any calls ($durNoCalls$), and the standard deviation of these durations ($durNoCalls_{std}$) were calculated to be used as social contact features. For these two active data feature sets, the performance of using the mean of all surveys inside the window or using the most recent survey was also tested.

Using the participant's location data, the most frequently visited location was determined and defined as the ``home" location. The number of times the participant was at the home location was calculated and used as a feature ($atHome$). For the second location feature, Haversine distances between all locations to the home location were summed ($distToHome$). Lastly, the area within a 2 km radius from home was defined as “zone-1”. The area outside of this radius was defined as  “zone-2”. The number of times the participant contributed from these two zones were calculated ($zone_{1}$, $zone_{2}$).

From the KCCQ data, two different feature sets were investigated. Firstly, the summation score ($KCCQ_{sum}$) was used as a feature. For the second set ($KCCQ_{all}$), each domain (physical limitation, symptom frequency, quality of life, and social limitation) was used separately.

\subsection{Machine learning models}
Logistic regression classifiers were trained to map the feature vector to the compensated or decompensated outcome. All the models were written in the Python 3 language, and the programming code was based on Scikit-learn \cite{sklearn_api}. Since each participant could contribute more than one event, we used leave-one-subject out cross-validation. The model was trained on the data from all participants except one held-out participant, and this participant's data was used as the test set. This process was repeated for each participant in the dataset.

Since the number of compensated and decompensated events were highly imbalanced, as seen in Table \ref{dataset}, the majority undersampling was performed on the training set before training the classifiers. During the majority undersampling, all participants from the minority class were used, and the same number of participants from the majority class were randomly selected. Sequential forward feature selection was used to select the three most informative features from each modality.

\begin{figure}[!htbp]
    \centering
    \includegraphics[width=0.40\textwidth]{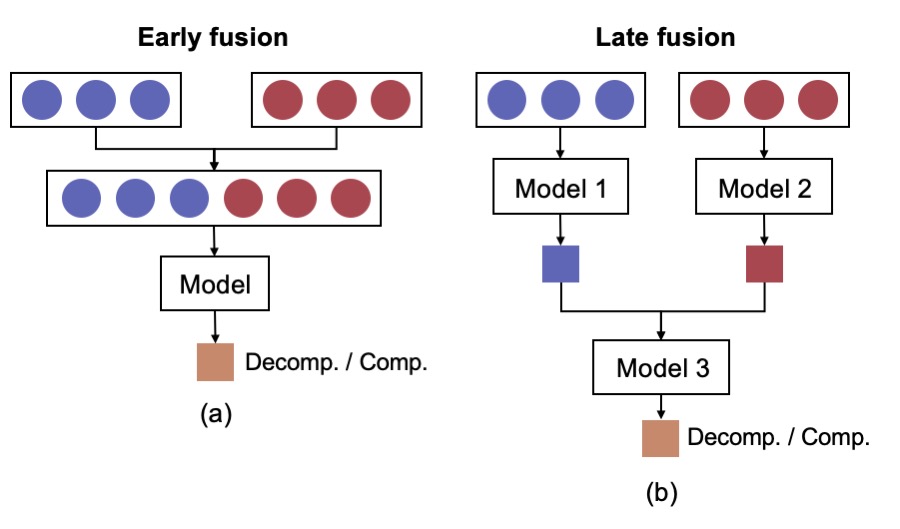}
    \caption{Modality fusion techniques. Purple and red colors indicate two different modalities. Figure (a) shows the early fusion approach, and figure (b) shows the late fusion of the modalities.}
    \label{fusionPlot}
\end{figure}

Early and late fusion approaches combined passive and active modalities and are shown in Fig. \ref{fusionPlot}. In the early fusion approach, extracted features were combined at the input level to create a single feature vector. Secondly, all single modality model's output probabilities were concatenated and used as input to another classifier for the late fusion approach. In the fusion models, the participants who contributed all data types were included in the analysis. 

To examine and interpret the features further, SHapley Additive exPlanation (SHAP) values for the early fusion model were calculated \cite{lundberg2017unified}. This framework is model agnostic, and SHAP values quantify the contribution and impact of each feature to the model. 

\subsection{Sedentary activity recognition}
\label{walkmodel}
To better interpret the results of the experiments, a smartphone-based sedentary activity recognition model using The Human Activity Recognition database was implemented \cite{anguita2013public}. This database consists of 30 participants doing daily-life activities such as walking or sitting. Walking activity categories (walking, walking-upstairs, walking-downstairs) and sedentary activity categories (sitting, standing, laying) were combined and used for training a binary sedentary activity classifier. Random forest, logistic regression, and decision tree classifiers were trained using the raw accelerometer mean, and standard deviation features derived from 2.56-second windows. A random forest classifier was selected since it achieved a 5-fold cross-validated accuracy of 0.99 on the training set. Then, the classifier was applied to the dataset to obtain walking or sedentary labels for each 2.56-sec window with 50\% overlap.

\section{Experimental Results}
\subsection{Single modality model results}
The cross-validation performance for each single modality model is shown on Table \ref{tab:passiveSingleModResults}. For these experiments, the window was set to 14 days before each clinical event. The number of unique participants and the number of clinical events changed according to the modality since the participants could stop contributing data. For the motion model, 23 participants contributed 28 decompensated events and 44 compensated events. For the social contact model, there were 21 participants with 27 decompensated events and 45 compensated events. Lastly, there were 18 participants with 13 decompensated events and 33 compensated events for the location model. Most selected features by the feature selection algorithm were $act_{mean}$, $act_{mode}$, and $act_{comp}$ for motion; $durCalls_{std}$, $durNoCalls_{std}$, and $durNoCalls$ for social contact; $zone_{1}$, $atHome$, and $distToHome$ for the location model.

\begin{table}[!htbp]
\centering
\caption{Passive data model results. `Combined' indicates a model that uses all passive modalities.} 
\label{tab:passiveSingleModResults}
\begin{tabular}{@{}cccccc@{}}
\toprule[\heavyrulewidth]\toprule[\heavyrulewidth]
\textbf{Modality} & \textbf{Acc.} & \textbf{AUC} & \textbf{AUCPr} & \textbf{PPV} & \textbf{TPR} \\ \midrule
\vspace{0.15cm}
Motion & 0.65 &  0.66 & \textbf{0.61} & 0.55 & 0.61 \\
\vspace{0.15cm}
Location & 0.61 & 0.58 & 0.39 & 0.33 & 0.38 \\ \vspace{0.15cm}
Social &  \textbf{0.68} & 0.66 & 0.54 & \textbf{0.57} & 0.59 \\
Combined & 0.65 & \textbf{0.69} & 0.55 & 0.48 & \textbf{0.77} \\
\bottomrule
\end{tabular}
\end{table}

Table \ref{tab:activeSingleModResults} provides the single modality results for the active data type, KCCQ survey. For two different feature sets ($KCCQ_{sum}$ and $KCCQ_{all}$), the table shows the performance metrics when the mean of all the questionnaires within the 14-day window was used and when the most recent questionnaire was used. For this active data type, 20 unique IDs contributed 23 decompensated events and 32 compensated events. Using the summary KCCQ score and taking the most recent questionnaire has resulted in the highest AUCPr score of 0.74.

\begin{table}[!htbp]
\centering
\caption{Active data single modality model results.}
\label{tab:activeSingleModResults}
\begin{tabular}{@{}cccccc@{}}
\toprule[\heavyrulewidth]\toprule[\heavyrulewidth]
\textbf{Modality} & \textbf{Acc.} & \textbf{AUC} & \textbf{AUCPr} & \textbf{PPV} & \textbf{TPR} \\ \midrule
\vspace{0.15cm}
\begin{tabular}[c]{@{}c@{}}Mean of window,\\ $KCCQ_{sum}$\end{tabular} & 0.62 & 0.76 & 0.64 & 0.54 & 0.65 \\
\vspace{0.15cm}
\begin{tabular}[c]{@{}c@{}}Mean of window,\\ $KCCQ_{all}$\end{tabular} & 0.67 & 0.68 & 0.58 & 0.59 & 0.70 \\
\vspace{0.15cm}
\begin{tabular}[c]{@{}c@{}}Most recent,\\ $KCCQ_{sum}$\end{tabular} & 0.67 & {\bf 0.77} & {\bf 0.74} & 0.59 & 0.70 \\
\begin{tabular}[c]{@{}c@{}}Most recent,\\ $KCCQ_{all}$\end{tabular} & {\bf 0.73} & 0.71 & 0.63 & {\bf 0.65} & {\bf 0.74} \\ \bottomrule
\end{tabular}
\end{table}

\subsection{Modality fusion model results}
\begin{figure}[htp!]
    \centering
    \includegraphics[width=0.40\textwidth]{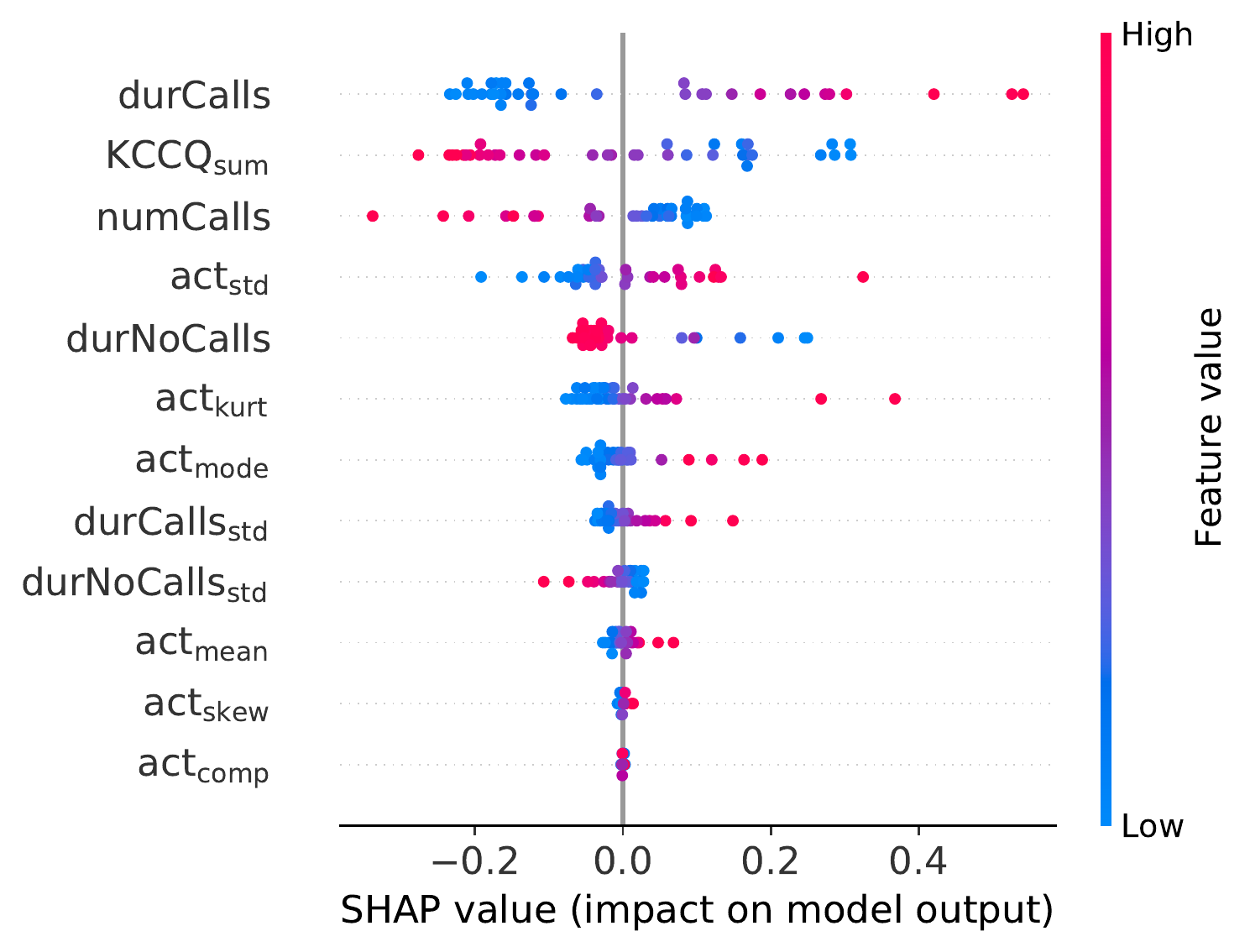}
    \caption{SHAP summary plot for the early fusion model. Features are sorted by their impact on the y-axis. Each point on the plot shows the Shapley value for one instance. Horizontal location shows the feature's effect for predicting positive class (decompensated) or negative class (compensated), and color indicates the feature value.}
    \label{shapPlot}
\end{figure}

In the fusion of KCCQ and motion data, 17 participants contributed data for both modalities, 21 decompensated events, and 26 compensated events. When three modalities were used (KCCQ, motion, social contact), 16 participants contributed 18 decompensated events and 21 compensated events. Lastly, when all data types were merged, there was data available for 12 participants, ten decompensated events, and 18 compensated events. The results for the early fusion models is shown in Table \ref{tab:earlyFusionResults} and in Table \ref{tab:lateFusionResults} for the late fusion models. The highest AUCPr of 0.80 was achieved when KCCQ and motion and social contact modalities were combined with late fusion. For early fusion models, using the same modalities resulted in an AUCPr of 0.74. The corresponding SHAP summary plot for the early fusion model is shown in Fig. \ref{shapPlot}.

\begin{table}[!htbp]
\centering
\caption{Results of early fusion models.}
\label{tab:earlyFusionResults}
\begin{tabular}{@{}cccccc@{}}
\toprule[\heavyrulewidth]\toprule[\heavyrulewidth]
\textbf{Modality} & \textbf{Acc.} & \textbf{AUC} & \textbf{AUCPr} & \textbf{PPV} & \textbf{TPR} \\ \midrule
\vspace{0.17cm}
KCCQ, motion & {\bf 0.77} & {\bf 0.82} & {\bf 0.76} & {\bf 0.73} & {\bf 0.76} \\
\vspace{0.15cm}
\begin{tabular}[c]{@{}c@{}}KCCQ, \\ motion, soc.\end{tabular} & 0.74 & 0.74 & 0.74 & 0.72 & 0.72 \\
\begin{tabular}[c]{@{}c@{}}KCCQ, motion, \\ soc., loc.\end{tabular} & 0.71 & 0.66 & 0.62 & 0.58 & 0.70 \\ \bottomrule
\end{tabular}
\end{table}

\begin{table}[!htbp]
\centering
\caption{Results of late fusion models.}
\label{tab:lateFusionResults}
\begin{tabular}{@{}ccccccc@{}}
\toprule[\heavyrulewidth]\toprule[\heavyrulewidth]
\textbf{Modality} & \textbf{Acc.} & \textbf{AUC} & \textbf{AUCPr} & \textbf{PPV} & \textbf{TPR} \\ \midrule
\vspace{0.17cm}
KCCQ, motion & 0.70 & 0.75 & 0.67 & 0.65 & 0.71 \\
\vspace{0.15cm}
\begin{tabular}[c]{@{}c@{}}KCCQ, \\ motion, soc.\end{tabular} & {\bf 0.77} & {\bf 0.82} & {\bf 0.80} & {\bf 0.70} & {\bf 0.89} \\
\begin{tabular}[c]{@{}c@{}}KCCQ, motion, \\ soc., loc.\end{tabular} & 0.64 & 0.79 & 0.67 & 0.50 & 0.90 \\ \bottomrule
\end{tabular}
\end{table}

\subsection{Time-to-event analysis}
\begin{figure}[!htbp]
    \centering
    \includegraphics[width=0.45\textwidth]{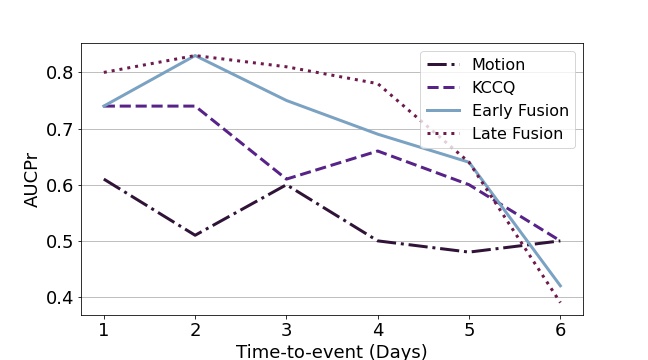}
    \caption{Performance change as the data window is shifted. x axis indicates the time-to-event. Early and late fusion models use KCCQ, motion, social contact modalities.}
    \label{timeToEvent}
\end{figure}

Using the best models in each category, how early the algorithm could predict the outcome (time-to-event analysis) was also investigated. Figure \ref{timeToEvent} illustrates the AUCPr of the models as the window was shifted. For all models, using the data up until the end of the day before the event resulted in the highest AUCPr. However, a similar performance was observed four days before decompensation for the late fusion model.

\subsection{Sedentary activity recognition results and in-depth analysis of motion data}
\begin{figure}[htp!]
    \centering
    \includegraphics[width=0.43\textwidth]{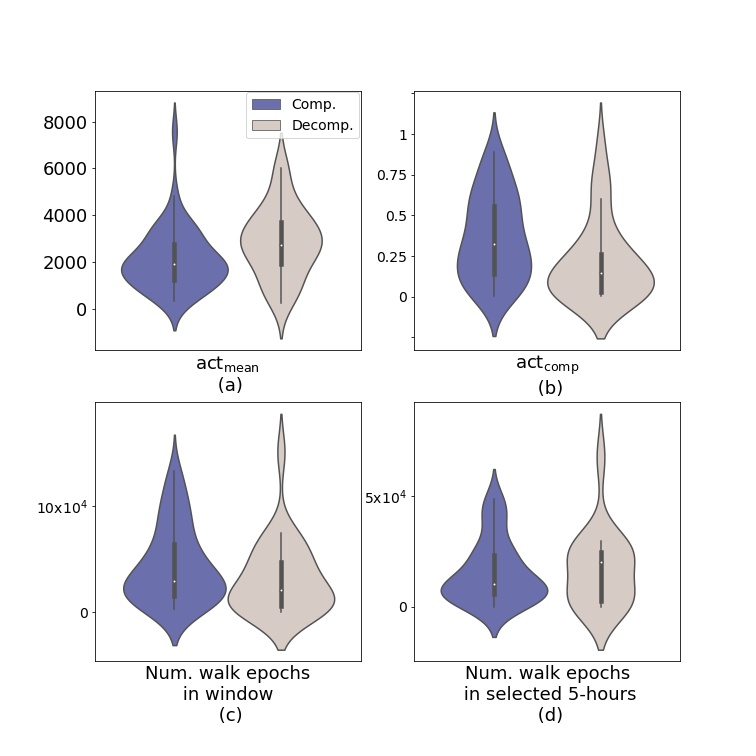}
    \caption{Violin plots of motion features $act_{mean}$ and $act_{comp}$, and walk epoch counts for the window and randomly selected 5-hours with over 50\% completeness.}
    \label{boxPlot}
\end{figure}

The purpose of the experiments in this section was to investigate the motion data in further detail. Fig. \ref{boxPlot} subplots (a) and (b) show the $act_{mean}$ and $act_{comp}$ features for the clinical event categories. Subplot (c) shows that the number of detected walk epochs was higher for the compensated windows. However, when 5-hours with 50\% maximum missingness was randomly sampled from each window, the difference between the classes was not significant, as shown in subplot (d).

\section{Discussion}
In this work, the features derived from data passively collected by a smartphone app were used for predicting decompensation events in a heart failure population. There were three passive data modalities (motion, location, and social interactions) and one active (the KCCQ). Combining the patient-reported KCCQ scores with the passive metrics resulted in the models with the best performance.

Firstly, next-day prediction algorithms were built using each modality separately. From the passive data sources, the motion data-based model achieved the highest AUCPr of 0.61. For a model based only on the responses of the KCCQ, using the summary of all domains and using the most recent score resulted in the best performance with an AUCPr of 0.74 (Table \ref{tab:activeSingleModResults}). 

Combining both passive and active data modalities achieved a superior performance compared to models based on passive or actively collected data alone (see Tables \ref{tab:earlyFusionResults} and \ref{tab:lateFusionResults}). The highest performing model combined KCCQ, motion, and social contact data. Using the late fusion approach achieved a 6\% higher AUCPr compared to early fusion when three modalities were used. Late fusion summarizes each modality and presents a lower-dimensional vector to the final classifier \cite{huang2020fusion}. Therefore, this method could reduce the chances of overfitting and addresses the curse of dimensionality when the sample size is small. The high true positive rate (0.89) and positive predictive value (0.70) of this model could indicate that the approach could potentially add clinical interventions into the framework and result in a low number of false alarms.

Figure \ref{shapPlot} illustrates the feature importance using the SHAP method. Duration and number of calls were among the most informative features, indicating that the dynamics of social interactions could be affected by the disease status. The SHAP summary plot also indicates that a higher duration but fewer calls result in a higher probability of HF decompensation for the model. Another important feature was the KCCQ summary value, and a lower value of this parameter gave rise to higher SHAP values. 

The SHAP plot also indicated that higher mean smartphone motion intensity resulted in a higher probability of HF, which was unexpected since HF limits daily physical activity and is often associated with fatigue. Figure \ref{boxPlot} shows that the mean activity was higher in the decompensated windows in subplot (a), but that the completeness of the data was much lower. The number of walk epochs inside the window was also calculated using the approach outline in Section \ref{walkmodel}, and the compensated windows had a slightly higher number of walk epochs. However, this was also affected by the imbalance of data completeness between the classes. When five hours of data with at least 50\% completeness was sampled to mitigate the effects of this imbalance, the difference between the classes was not significant.  In a previous study, Duncan \etal have shown that steps measured by a smartphone and a wearable differed a mean bias of 21.5\%, and hypothesize that this could result from the behavior of the participants (i.e., not carrying the phone on short walking breaks, carry location for the phone) \cite{duncan2018walk}. Similarly, our results show that the smartphone's motion data does not measure the physical effort but that it reflects patterns of behavior, including phone utilization and body movements.

When different time-to-event horizons were tested, a general trend of lower performance for longer future predictions was observed, as expected, since symptoms are likely to become more pronounced closer to the event. However, predictions two days ahead were actually better than one day, and performance four days ahead was almost as good as one day before the event. This indicates that one-day, two-day, and four-day models could be run simultaneously to identify short- and medium-term risks and result in different levels of intervention. Changes in performance will be affected by the levels of missingness as the event is approached, as well as the intrinsic behaviors, which may explain the performance of the two-day window.

There are two key limitations of the study presented in this article. Firstly, when the data were missing, the app did not indicate whether this resulted from the participant closing the app voluntarily or if it resulted from the smartphone battery running out. These behaviors have different etiologies, which may be related to impending decompensation in different ways. For example, closing the app may indicate being tired, whereas a  battery running out of charge may indicate apathy connected with depression. If an additional label is collected for missing sections, it could be used to learn other behavioral patterns. Secondly, even though each participant contributed many days, the study's sample size was relatively small (N=28 participants), and therefore, the methods should be further validated in a larger cohort.

\section{Conclusion}
A smartphone-based approach for monitoring HF patients non-invasively has been proposed, which may provide adequate performance for clinical interventions. The proposed app-based framework collects motion, social contact, location data and administers clinically-validated surveys to monitor HF severity of the participants. We hypothesize that due to the ubiquity of smartphones and the ease of scalability of the framework, our method will facilitate monitoring large populations at a low cost. In future work, the feasibility of combining the proposed method with clinical interventions (such as teleconsults and drug dose modification) will be investigated to measure the potential impact of the framework described in this work. 

\section{Acknowledgments}
The authors wish to acknowledge the support of the National Science Foundation Award 1636933, ``BD Spokes: SPOKE: SOUTH: Large-Scale Medical Informatics for Patient Care Coordination and Engagement'', NIH/NHLBI award K23 127251, the Georgia Research Alliance, and the National Center for Advancing Translational Sciences of the National Institutes of Health under Award Number UL1TR002378. 

\section{References}
\printbibliography[heading=none]

\end{document}